\def\annp #1 #2 #3 {{\sl Ann.\ Phys.\ (N.Y.)} {\bf #1}, #2 (#3)}
\def\jpa #1 #2 #3 {{\sl J. Phys.\ A} {\bf #1}, #2 (#3)}
\def\pra #1 #2 #3 {{\sl Phys.\ Rev.\ A} {\bf #1}, #2 (#3)}
\def\pre #1 #2 #3 {{\sl Phys.\ Rev.\ E} {\bf #1}, #2 (#3)}
\def\prl #1 #2 #3 {{\sl Phys.\ Rev.\ Lett.} {\bf #1}, #2 (#3)}
\def\jsp #1 #2 #3 {{\sl J. Stat.\ Phys.} {\bf #1}, #2 (#3)}
\def\jmp #1 #2 #3 {{\sl J. Math.\ Phys.} {\bf #1}, #2 (#3)}
\def\zpb #1 #2 #3 {{\sl Z. Phys.\ B} {\bf #1}, #2 (#3)}
\def\jcp #1 #2 #3 {{\sl J.\ Chem.\ Phys.} {\bf #1}, #2 (#3)}
\def\jpc #1 #2 #3 {{\sl J.\ Phys.\ Chem.} {\bf #1}, #2 (#3)}
\def\zpc #1 #2 #3 {{\sl Z.\ Phys.\ Chem.} {\bf #1}, #2 (#3)}
\def\anp #1 #2 #3 {{\sl Ann.\ Prob.\ } {\bf #1}, #2 (#3)}
\def\cp #1 #2 #3 {{\sl Chem.\ Phys.\ } {\bf #1}, #2 (#3)}
\def\annp #1 #2 #3 {{\sl Ann.\ Phys.\ (N.Y.)} {\bf #1}, #2 (#3)}
\def\mplb #1 #2 #3 {{\sl Mod.\ Phys.\ Lett.\ B} {\bf #1}, #2 (#3)}
\def\jetp #1 #2 #3 {{\sl Sov.\ Phys.\ JETP} {\bf #1}, #2 (#3)}
\def\ie{{\it i.\ e.}}\def\etc{{\it etc.}}

\def\td#1#2{{d #1\over d #2}}      %total derivative
\def\t2d#1#2{{d^2 #1\over d #2^2}} %second total derivative
\def\p2d#1#2{{\partial^2 #1\over\partial #2^2}} %second partial derivative
\def\gtwid{\mathrel{\raise.3ex\hbox{$>$\kern-.75em\lower1ex\hbox{$\sim$}}}}
\def\ltwid{\mathrel{\raise.3ex\hbox{$<$\kern-.75em\lower1ex\hbox{$\sim$}}}}

\def\nbt{{n_B(t)}}
\def\nit{{n_I(t)}}
\def\as#1#2{{#1\sim t^{-#2}}}
\def\db{{\bar d}}
\def\kbb{{k_{BB}^d}}
\def\kbi{{k_{BI}^d}}
\def\kbib{{k_{BI}^\db}}

\def\ai{{\alpha_I}}
\def\ad{{\alpha(d,\db\,)}}
\def\bd{{\beta(d,\db\,)}}
\documentstyle[12pt]{article}
\begin{document}

\begin{titlepage}
\centerline{\bf Reaction Kinetics of Clustered Impurities}
\bigskip
\centerline{\bf E.~Ben-Naim}\smallskip
\centerline{The James Franck Institute, The University of Chicago}
\centerline{Chicago, IL 60637}
\vskip .65in
\centerline{ABSTRACT}
{\noindent We study the density of clustered immobile reactants in the
diffusion-controlled single species annihilation. An initial state
in which these impurities occupy a subspace of codimension $\db$ leads to
a substantial enhancement of their survival probability.  The Smoluchowski
rate theory suggests that the codimensionality plays a crucial role in
determining the long time behavior.  The system undergoes a transition
at $\db=2$. For $\db<2$, a finite fraction of the impurities survive:
$\nit=n_I(\infty)+{\rm const}\times\log(t)/\sqrt{t}$ for $d=2$ and
$\nit=n_I(\infty)+{\rm const}/\sqrt{t}$ for $d>2$.  Above this
critical codimension, $\db\ge 2$, the subspace decays indefinitely.
At the critical codimension, inverse logarithmic decay occurs,
$\nit\sim \log(t)^{-\alpha(d,\db)}$.  Above the critical codimension, the
decay is algebraic $\as \nit {\alpha(d,\db)}$. In general, the exponents
governing the long time behavior depend on the dimension as well as
the codimension.}
\vskip.3in
\noindent P.A.C.S. Numbers: 02.50.-r, 05.40+j, 82.20.-w, 82.20.Wt
\end{titlepage}

\centerline{\bf I. Introduction}\smallskip

The kinetics of diffusion-controlled chemical reactions have attracted
much interest recently. For simple homogeneous reaction
processes, substantial theoretical knowledge is available
\cite{Torney-83,Racz-85,Kang-85,Lushnikov-86,Spouge-88,ben-Avraham-90}.
For single species bimolecular reactions it is well known that for
$d\le 2$, spatial correlations between reacting particles are
important in  the long time limit, while for $d>2$
reactants are effectively transparent, and a universal decay of the
density occurs.  A number of reaction processes, such as the
annihilation processes, $A+A\to 0$ \cite{Kang-85} and the aggregation
process, $A+A\to A$ \cite{ben-Avraham-90} belong to this universality
class.

A recent generalization of this process to heterogeneous situations
was shown to exhibit a rich array of asymptotic behavior. When the
reactants have a polydisperse distribution of diffusion coefficients,
non-universal decay kinetics may occur. Such a process is well suited
for describing reactions that involve particles with different masses.
The case in which a small number of particles move according to one
diffusion coefficient and the bulk according to another is especially
interesting because of  its simplicity. The survival probability of a
single ``impurity'' particle immersed in a background of identical
particles, $\nit$, depends in a nonuniversal fashion on the
diffusivity of the impurity. In one dimension, the results are
especially intriguing. While for the aggregation process, a mapping to
a random walk in three dimension enables an exact solution
\cite{Fisher-84}, for the annihilation process only an approximate
theory is available \cite{Krapivsky-94}.  The survival probability of
an immobile impurity particle is equivalent to the fraction of
unvisited sites by annihilating random walkers
\cite{Cardy-94}. Additionally, as the annihilation process is
equivalent to the $T=0$ Ising model with Glauber dynamics
\cite{Glauber-63}, $\nit$ equals the fraction of ``cold'' spins,
\ie, the number of spins that did not flip up to time
$t$. Numerically, both time-dependent and finite size simulations
\cite{Krapivsky-94,Derrida-94,Stauffer-94}, as well as time series
studies \cite{Krapivsky-94,Ben-Naim-93} suggest that the impurity
survival probability follows a non-trivial decay $\as \nit \ai$, with
$\ai\cong0.37$, while the corresponding theoretical exponent for the
aggregation case is $\ai=1$. In addition, for trimolecular
annihilation, $A+A+A\to 0$ it was found that the impurity decays
faster than a power-law but slower than an exponential $\nit\sim
\exp\big(-\log(t)^{3/2}\big)$ \cite{Cardy-94}.  Hence, in one-dimension,
the impurity decay kinetics are highly sensitive to microscopic
details of the reaction process.

In this study, we present a generalization of the isolated impurity
problem.  We examine the collective effects of ``trapped'' or
equivalently immobile impurities by considering the initial condition
in which the immobiles are clustered. For simplicity, the immobiles
occupy a subspace of dimension $d_I$ embedded in a $d$-dimensional
space.  For convenience we introduce the codimensionality
$\db=d-d_I$. Note that the case $\db=d$ corresponds to the isolated
impurity problem.  We are primarily interested in the relevance of the
subspace, if any, to the asymptotic behavior of the survival
probability.  In the long time limit, neighboring immobile reactants
``shield'' each other, and as a result, the immobile reactants decay
significantly slower than the background. The subspace becomes more
pronounced and consequently, the subspace interface acts as an
absorbing boundary to the background. The geometry of the system
reduces to a $\db$-dimensional one.  By applying the Smoluchowski rate
theory to the background in dimension $d$ and to the impurity subspace
in dimension $\db$, theoretical prediction concerning the asymptotic
form of the survival probability are made possible. The subspace
undergoes a survival-extinction transition at $\db=\db_c=2$. Below
this critical codimension, a finite fraction of the immobiles survive,
while at a higher codimension they decay forever. For $\db=1$, the
approach to the final density is an algebraic one
$\nit-n_I(\infty)\sim 1/\sqrt{t}$ for $d>2$, with a logarithmic
correction at $d=2$, $\nit-n_I(\infty)\sim\log(t)/\sqrt{t}$.  At the
critical codimension, $\db=2$, an unusual logarithmic decay occurs
$\nit\sim\log(t)^{-\ad}$, while for $\db>2$ a dimension dependent decay
$\as \nit {\ad}$ occurs as well.

\medskip\centerline{\bf II. The Isolated Impurity Problem}\smallskip

In this section we review the rate equation theory and apply it to the
isolated impurity problem. This approach truncates the infinite
hierarchy of equations describing the density at the first order.
Above the critical dimension, spatial fluctuations are asymptotically
irrelevant and thus, this theory is exact. However, this theory can
be extended to arbitrary dimension, and is especially
attractive due to its simplicity.

In the lattice version of the homogeneous single species annihilation
process, particles hop independently with rate $D_B$ to any one of
their nearest-neighbor sites.  An attempt to land on an occupied site
results in the removal of both particles from the system. Hence, the
density, $\nbt$, obeys the following rate equation,
$dn_B/dt\propto -n_{BB}$, where $n_{BB}$ is the density of pairs of
neighboring particles. This approach leads to an infinite
hierarchy of rate equations and is of limited practical
use. Alternatively, one assumes that the annihilation rate is
proportional to the flux experienced by a particle, $dn_B/dt\propto
-jn_B$ \cite{Smoluchowski-17}. This flux can be evaluated by placing
an absorbing particle in a diffusing background of density $n_B$.
Since $j$ is proportional to $n_B$, the density is described by the
following rate equation,
\begin{equation}
\td {n_B} t=-\kbb n_B^2,
\label{Eq-1}
\end{equation}
with $\kbb$ being the reaction rate. Above the critical dimension
$d_c=2$ the reaction rate is constant, and the above theory is exact.
Otherwise, the reaction rate is time dependent since the target
particle is surrounded by a depletion zone the size of the diffusion
length $\sqrt{Dt}$. In appendix I, we detail a heuristic derivation
of the reaction rate in arbitrary dimension using the quasistatic
approximation.  In general, $k_{IJ}^d$, the
reaction rate between particles of diffusivities $D_I$ and $D_J$ has the
following long time form,
\begin{equation}
k_{IJ}^d\simeq\cases {A_1\sqrt{(D_I+D_J)/t}&$d=1$;\cr
                   A_2 (D_I+D_J)/\log\left((D_I+D_J)t\right)&$d=2$;\cr
                   A_d (D_I+D_J)&$d>2$.\cr}
\label{Eq-2}
\end{equation}
Note that both $k_{IJ}^d$ and $D_I$ are rates and that $A_d$ is a
dimensionless prefactor. By introducing a modified time variable
\begin{equation}
z(t)=\int_0^t \kbb(t')dt',
\label{Eq-3}
\end{equation}
the rate equation simplifies,
$dn_B/dz=-n_B^2$. The asymptotic solution to this equation is
$n_B\sim 1/z$. Evaluating the modified time variable, $z$, we arrive
at the following asymptotic behaviors of the homogeneous
single-species annihilation process,
\begin{equation}
\nbt\sim\cases    {1/\sqrt{t}&$d=1$;\cr
                   \log(t)/t &$d=2$;\cr
                   1/t       &$d>2$.\cr}
\label{Eq-4}
\end{equation}
Interestingly, these results are asymptotically exact
\cite{Lushnikov-86,Spouge-88}, despite the assumptions
involved with the rate theory. Moreover, the same results are obtained
for the aggregation process $A+A\to A$ as well.  In lower dimensions,
the reaction proceeds with a slower rate, since particles are
effectively repelling each other. On the other hand, in high
dimensions spatial correlations are practically irrelevant and the
density decays faster.  The critical dimension is characterized by
typical logarithmic corrections.

In the impurity problem, the survival probability
 of a single impurity with diffusivity $D_I$, $\nit$, is investigated.
Eq.~(\ref{Eq-1}) can be generalized to situations with more
than one species
\begin{equation}
\td {n_I} t=-\kbi n_B n_I,
\label{Eq-5}
\end{equation}
with $\kbi$ given by Eq.~(\ref{Eq-2}).  Again, it is useful to rewrite
the rate equation in terms of the modified time variable $z$ defined
by Eq.~(\ref{Eq-3}), \hbox{$dn_I/dz=-(\kbi/\kbb)n_Bn_I$}.  Note that
in the long time limit the rate ratio approaches a constant that
depends only on the diffusivity ratio.  By substituting the asymptotic
form of the background density $1/z$, one finds a purely algebraic
dependence of the impurity density $n_I\sim z^{\kbi/\kbb}$ for
$d\ne2$.  For the case $d=2$ the leading asymptotic
correction to the rate ratio is important and a
slightly more detailed calculation is needed.  Using the
aforementioned forms of $z$ and $k_{IJ}^d$, we find the following
behavior for the impurity density,
\begin{equation}
\nit\sim\cases{t^{-\sqrt{(D_B+D_I)/(8D_B)}}&$d=1$;\cr
            t^{-(D_B+D_I)/2D_B}\log(t)^{\gamma}&$d=2$;\cr
            t^{-(D_B+D_I)/2D_B} &$d\ge2$.\cr}
\label{Eq-6}
\end{equation}
Hence, the diffusivity ratio, $r=(D_I+D_B)/2D_B$, governs the long
time kinetics.  The logarithmic correction for the case $d=2$ is
characterized by $\gamma=r(1-\log r)$ \cite{Krapivsky-94a}.  For the
case $D_I=0$, the decay exponent equals $1/\sqrt{8}\cong0.353$ when
$d=1$, and $1/2$ when $d>2$.
While the latter value is exact, the former is only
approximate
\cite{Krapivsky-94}.  Interestingly, this approximate value is quite
close to the observed numerical value $0.37$.  However, as this
approximation is uncontrolled, its accuracy can widely vary, and for
the exactly soluble aggregation process, the discrepancy in the
exponents is much larger.  To summarize, both in the supercritical
regime and in the subcritical regime, the decay of the impurities
depends on the diffusivity ratio in a nongeneric fashion.

\medskip\centerline{\bf III. Kinetics of Clustered Impurities}\smallskip

\begin{figure}[t]
\begin{center}
\begin{picture}(200,120)(40,640)
\thicklines
\put( 60,740){\circle{15}}
\put( 60,700){\circle{15}}
\put( 60,660){\circle{15}}
\put(100,660){\circle{15}}
\put(100,700){\circle{15}}
\put(100,740){\circle{15}}
\put(220,740){\circle{15}}
\put(220,700){\circle{15}}
\put(220,660){\circle{15}}
\put(180,660){\circle{15}}
\put(180,700){\circle{15}}
\put(180,740){\circle{15}}
\put(140,660){\circle*{15}}
\put(140,700){\circle*{15}}
\put(140,740){\circle*{15}}
\put(240,760){\line( 0,-1){120}}
\put(200,760){\line( 0,-1){120}}
\put(160,760){\line( 0,-1){120}}
\put(120,760){\line( 0,-1){120}}
\put( 80,760){\line( 0,-1){120}}
\put( 40,760){\line( 0,-1){120}}
\put( 40,640){\line( 1, 0){200}}
\put( 40,680){\line( 1, 0){200}}
\put( 40,720){\line( 1, 0){200}}
\put( 40,760){\line( 1, 0){200}}
\end{picture}
\end{center}
\caption[Illustration of the impurity problem]{A line of
immobile impurities (bullets) in a two dimensional background (circles).}
\label{Fig-1}
\end{figure}
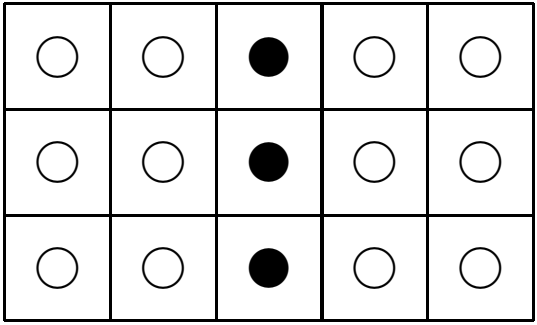

The following questions arise naturally from the above theory: Can the
presence of neighboring impurities increase the survival probability
of an impurity?  If yes, to what extent?  To answer these questions we
start with an initial configuration where impurities occupy a subspace
of dimension $d_I$.  The codimension $\db$  can be conveniently
defined as $\db=d-d_I$.  For example, when $\db=1$, the impurities
initially occupy a line in 2D, a plane in 3D, \etc\  Figure
\ref{Fig-1} illustrates the initial configuration for the case $\db=1$
in two spatial dimensions.  Note also that the single impurity problem
corresponds to the special case $\db=d$.  To preserve the geometrical
properties of the impurity subspace as the reaction evolves, we
restrict our attention to the case $D_I=0$.

Let us consider a line of impurities in 2D.  The presence of nearby
static particles can only increase the survival probability of an
impurity, and thus, the survival probability is bounded by $\as {n_I}
{1/2}$, the corresponding result for the single impurity case for
$d\ge2$. However, the background decay $\as {n_B} 1$ is stronger and
consequently, the subspace becomes more and more pronounced in the long
time limit. Hence, the impurity-background interface is equivalent to an
absorbing boundary for the mobile reactants. This absorber is not a
perfect one since not all sites are occupied with impurities.
However, it is well known that in the long time limit a partial
absorber is equivalent to a perfect one. A depletion layer of width
$\sqrt{Dt}$ develops around the subspace and the background density
profile is strongly suppressed in this depletion zone.  As a result,
the geometry of the system is drastically altered.  A line in 2D
reduces to a one-dimensional geometry.  In other words, the
codimension becomes the relevant dimension.  This simple conclusion
has a striking effect on the long time kinetics of the impurities.

One possible way to tackle this problem is to describe this spatial
inhomogeneity by the reaction-diffusion equation with proper
boundary conditions accounting for the impurity interface. However,
the leading term in the reaction-diffusion equation is the diffusion
term and this approach is equivalent in the long time limit to
the quasistatic approximation, or namely, the Smolechowsky rate
theory. This theory, appealing in its simplicity, leads to
new and interesting behaviors for the clustered impurity problem. As
discussed above, the relevant dimension is the codimension. Thus, we
apply the $\db$-dimensional rate equation to the impurities,
\begin{equation}
\td {n_I} t=-\kbib n_B n_I,
\label{Eq-7}
\end{equation}
with the reaction rates of Eq.~(\ref{Eq-2}).  The equation describing
the background is left unchanged, and one simply substitutes the
density from Eq.~(\ref{Eq-4}). In contrast with the previous analysis,
introduction of a modified time variable would not simplify the
algebra, since the equations involve different dimensions and
consequently, different intrinsic time scales.  Instead, the
impurity survival probability, $n_I(t)/n_I(0)$, is obtained by an
integration of the above equation,
\begin{equation}
\nit/n_I(0)=\exp\left(-\int_0^t dt' \kbib(t') n_B(t')\right).
\label{Eq-8}
\end{equation}
Since the case $\db=d$ reduces to the single impurity case (see
Eq.~(\ref{Eq-6})), we concentrate on the case $\db<d$ only.  By
substituting the proper values for the $d$-dimensional background density
and the $\db$-dimensional reaction rate into Eq.~(\ref{Eq-8}), we find
the following asymptotic impurity densities,
\begin{equation}
\nit\sim
\cases{
n_I(\infty)+{\rm const}\times\log(t)/\sqrt{t}& $\db=1$ and $d=2$;\cr
n_I(\infty)+{\rm const}/\sqrt{t}& $\db=1$ and $d>2$;\cr
\log(t)^{-A_{\db}/2A_{d}}&$\db=2$ and $d>\db$;\cr
t^{-A_{\db}/2A_{d}}&$\db>2$ and $d>\db$.\cr
}
\label{Eq-9}
\end{equation}
This rich behavior follows directly from the annihilation rate, \ie,
the integrand in Eq.~(\ref{Eq-8}). If $\kbib(t) n_B(t)$ decays faster
than $1/t$, the integral remains finite in the infinite time limit,
and a finite fraction of the impurities survive the annihilation
process.  Otherwise, the subspace vanishes. For $\db<2$, this
integrand decays faster than $1/t$, while for $\db\ge2$, the integrand
is dominated by $1/t$. Consequently, the system exhibits a
survival-extinction transition at $\db=2$.  Below this critical
codimension, a fraction of the impurities survive, while they decay
indefinitely at a higher codimension.

Interestingly, the approach to the final density,
$\nit-n_I(\infty)\propto\log(t)/\sqrt{t}$ for a line in 2D ($\db=1$)
is reminiscent of the single impurity decay in 2D. These cases differ
in that $n_I(\infty)$ does not vanish for the clustered case.  They
also differ in their logarithmic correction.  The critical case is
characterized by inverse logarithmic decay since the annihilation rate
is proportional to
$1/\left(t\log(t)\right)=d\log\left(\log(t)\right)/dt$. Both the
critical case and the supercritical regime follow a power-law with the
exponent, $\ad=A_{\db}/2A_{d}$. Thus, both decays depend on the
dimension as well as the codimension.  Only in the extreme case,
$\db=d$, corresponding to the single impurity case, $A_d$ cancels out
and a universal decay is expected, $\as {n_I} {1/2}$. Generally, a
detailed calculation for the prefactors is necessary in order to find
the various decay exponents.  Since the decay in the clustered case
$\db<d$ is slower than in the isolated case $\db=d$, we learn that the
prefactor $A_d$ is an {\it increasing} function of the spatial
dimension $d$. This observation is consistent with the fact that
the initial reaction rate is given by
$k_{IJ}^d(t=0)=2(D_I+D_J)z_d$, with $z_d$ the number of neighboring
sites in $d$-dimensions. Indeed, $z_d$ is an increasing function of
$d$. Above the critical dimension $d=2$, the effective reaction rate
depends weakly on time.  Hence, the assumption
$A_d\propto z_d$ leads to an approximate value for the decay exponent
$\ad\cong z_\db/2z_d$. For a simple square lattice one has the
$z_d=2d$ or $\ad\cong \db/2d$. Note also that this approximation
improves as the dimension and the codimension increase.

The above results can be easily generalized to arbitrary dimensions.
Such a generalization is nontrivial only when the temporal
nature of the reaction rate is dimension dependent, or namely, below
the critical dimension. Using the reaction rates of Eq.~(A3), we
evaluate the immobile impurity densities,
\begin{equation}
\nit\sim
\cases{
t^{-d/2^{1+d/2}}& $\db<2$ and $d=\db$;\cr
n_I(\infty)+{\rm const}\times t^{-(d-\db\,)/2}& $\db<2$ and $\db<d<2$;\cr
n_I(\infty)+{\rm const}\times\log(t)t^{-(2-\db\,)/2}& $\db<2$ and $d=2$;\cr
n_I(\infty)+{\rm const}\times t^{-(2-\db\,)/2}& $\db<2$ and $d>2$;\cr
t^{-1/2}\log(t)^{(1+\log 2)/2}& $\db=2$ and $d=2$\cr
\log(t)^{-A_{\db}/2A_{d}}&$\db=2$ and $d>\db$;\cr
t^{-A_{\db}/2A_{d}}&$\db>2$ and $d\ge\db$.\cr
}
\label{Eq-10}
\end{equation}
To summarize, the impurity subspace survives only when $\db<2$ and
$d>\db$. The behavior for $d<2$ is influenced by the background
density behavior $\as {n_B} {d/2}$. As a result, the approach towards
the limiting density is algebraic with a vanishing decay exponent
$(d-\db\,)/2=d_I/2$, for the case $\db\ltwid d<2$.  It will be interesting
to see how well the above results compare with Renormalization Group
studies in the vicinity of the critical codimension $\db=2-\epsilon$.

For completeness, we briefly discuss the early time behavior of the
system. We focus on the case where all lattice sites are initially
occupied, such that $n_B(0)=n_I(0)=1$.  Following the above
discussion, the initial background-background reaction rate is given
by $\kbb(t=0)\cong2Dz_d$. From Eq.~(\ref{Eq-1}), the background
density is found
\begin{equation}
n_B(t)\cong \left(1+2Dz_d t\right)^{-1}\qquad t\to 0.
\label{Eq-11}
\end{equation}
On the other hand, only interfacial sites
contribute to annihilation of impurities, and as a result
$\kbi=D(z_d-z_{d-\db})$. By substituting this rate and the early time
background density into Eq.~(\ref{Eq-5}), the impurity density is
calculated in the early time regime,
\begin{equation}
n_I(t)\cong \left(1+2Dz_d t\right)^{\bd}\qquad t\to 0.
\label{Eq-12}
\end{equation}
The above exponent, $\bd$, equals the reaction rate ratio
$\bd=(z_d-z_{d-\db})/2z_d$.  This exponent should not be regarded as
an asymptotic one, since it is relevant only for a short time.
In addition to the dimension dependence, the early time behavior
depends on the lattice structure as well. As the reaction process
evolves, such details become irrelevant, and the general asymptotic
behavior is recovered. Note that there is no sign of a critical
codimensionality in the early stages since the system is still
$d$-dimensional. After waiting a sufficiently long time, the geometry
changes, and the codimension governs the kinetics.

\medskip\centerline{\bf IV. Conclusions}\smallskip

We have studied the kinetic behavior of simple subspaces of immobile
reactants in the annihilation process $A+A\to0$. Asymptotically, such
an inhomogeneity leads to a change in the geometry of the system, and
the codimension becomes the relevant parameter.  The Smolechowsky
theory shows that a transition from survival to extinction takes place
at $\db=2$. For $\db<2$, and $d>\db$, a finite fraction of the
impurities survive, while for $\db\ge2$ the impurity subspace
eventually vanishes. Furthermore, the asymptotic behavior of the
impurity density depends on the dimension as well as the codimension.

This study suggests that the rate theory can be extended to
heterogeneous situations. Moreover, a time-dependent reaction equation
is equivalent in the long time limit to the detailed
reaction-diffusion equation.  The success of this theory is remarkable
especially considering its simplicity.  It will be interesting to
apply the same mechanism to more complicated processes, such as
multispecies reactions.

\clearpage
\medskip\centerline{\bf Acknowledgments}\smallskip

I am thankful to P.~Krapivsky and S.~Redner for useful discussions.
This work was supported in part by the MRSEC Program of the
National Science Foundation under Award Number DMR-9400379 and
by NSF under Award Number 92-08527.

\medskip\centerline{\bf Appendix A. The Quasistatic Approximation}\smallskip

The reaction rate, $k_{IJ}^d$, can be found
by evaluating the flux $j$ experienced by an absorbing test particle
of radius $R$ due to a diffusing background of concentration
$c_0$. Hence, the diffusion equation $\partial c/\partial t=D\nabla c$
is solved under the initial conditions \hbox{$c|_{t=0}=c_0$} and the
absorbing boundary condition $c(r)|_{r=R}=0$.  Although an exact
solution can be obtained, we present a simpler approximate technique.
Since the length scale governing the diffusion process is
$\sqrt{Dt}$, one assumes that the absorber does not affect the background
density at distances larger than $\sqrt{Dt}$. Inside this depletion zone
the Laplacian term dominates the spherically symmetric diffusion equation,
\renewcommand{\theequation}{A1}
\begin{equation}
Dr^{1-d}{\partial\over\partial r}r^{d-1}{\partial c(r)\over\partial r}=0,
\qquad R<r<\sqrt{Dt}.
\end{equation}
The quasistatic approximation imposes an additional time-dependent
boundary condition $c|_{r=\sqrt{Dt}}=c_0$ \cite{Redner-90}. The
concentration profile is readily obtained for $R<r<\sqrt{Dt}$,
\renewcommand{\theequation}{A2}
\begin{equation}
c(r,t)\simeq\cases{
c_0\left((r/R)^{2-d}-1\right)\Big/\left((\sqrt{Dt}/R)^{2-d}-1\right)&$d<2$;\cr
c_0\log(r/R)\big/\log(\sqrt{Dt}/R)&$d=2$;\cr
c_0\left(1-(r/R)^{2-d}\right)\Big/\left(1-(\sqrt{Dt}/R)^{2-d}\right)&$d>2$.\cr
}
\end{equation}
Above the critical dimension $d_c=2$, $c(r,t)$ approaches a limiting
density profile. The reaction rate is given by the
calculating the total flux seen by the test particle $j=DS_dR^{d-1}
\partial c/\partial r$, with $S_d$ the surface
area of the $d$-dimensional unit sphere.  We quote the leading
asymptotic term of the reaction rate, $k^d=j/c_0$,
\renewcommand{\theequation}{A3}
\begin{equation}
k\propto\cases{
D^{d/2}t^{d/2-1}&$d<2$;\cr
D/\log(Dt)&$d=2$;\cr
DR^{d-2}&$d>2$.\cr
}
\end{equation}
In case where the target diffusivity $D_T$ is nonzero, the effective
diffusion constant is \hbox{$D+D_T$}. We derived the continuum rates,
however, the lattice counterparts can be conveniently obtained by
setting $R\equiv 1$.

\listoffigures

\end{document}